# The Probability of Food Security:
# A new longitudinal data set using the Panel Study of Income Dynamics

Seungmin Lee*, John Hoddinott**, Christopher B. Barrett** and Matthew P. Rabbitt***

July 2025 revision

**Abstract:** The study of food security dynamics in the U.S. has long been impeded by the lack of extended longitudinal observations of the same households or individuals. This paper applies a newly-introduced household-level food security measure – the probability of food security (PFS) – to 26 waves of Panel Study of Income Dynamics (PSID) data, spanning 1979-2019, to generate a data product we describe and make newly available to the research community. We detail the construction of this unprecedentedly long food security panel data series in PSID data. Finally, we estimate key subpopulation- and national-level food security dynamics identifiable over the 40-year (1979-2019) period spanning multiple recessions and federal nutrition assistance policy changes, including disaggregated dynamics based on geography, race, sex, and educational attainment.

\* University of Notre Dame

\*\* Cornell University

\*\*\* Economic Research Service, U.S. Department of Agriculture

*Correspondence to be sent to: Seungmin Lee, Keough School of Global Affairs, University of Notre Dame, Notre Dame, USA. slee76@nd.edu*

*Date of submission: July 26, 2025*

Acknowledgements: We thank an anonymous reviewer and the editor, Craig Gundersen, for helpful comments on an earlier draft. This work was supported by USDA-Economic Research Service cooperative agreement 58-4000-1-0094 and by the intramural research program of the U.S. Department of Agriculture, National Institute of Food and Agriculture, Hatch grant 1023665. The collection of data used in this study was partly supported by the National Institutes of Health under grant number R01 HD069609 and R01 AG040213, and the National Science Foundation under award numbers SES 1157698 and 1623684. The findings and conclusions in this publication are those of the authors and should not be construed to represent any official USDA or U.S. Government determination or policy.

1. **Introduction**

Ensuring that all Americans enjoy sufficient access to safe, nutritious food - often labeled 'food security' - is one of the United States Department of Agriculture (USDA)'s strategic goals (USDA 2022). Although food insecurity is strongly associated with poverty, those two deprivations are not synonymous; many households with incomes above the poverty line are food insecure, and many poor households are food secure (Gundersen et al. 2011; Gundersen and Ribar 2011, Rabbitt et al. 2024). The United States (US) and other governments therefore set food security as an explicit goal separate from that of poverty reduction (Barrett 2002; Alderman et al. 2017). Assessing the effectiveness of such policies that address food insecurity requires measures that capture its prevalence, severity, and persistence.

In the United States, the prevalence and severity of food insecurity are calculated using the Food Security Scale Score (FSSS). The FSSS has been used by the USDA, Economic Research Service (ERS) since 1995 to generate these official annual estimates based on data collected in the prior year's Food Security Supplement to the Current Population Survey (CPS-FSS). The CPS-FSS includes a Household Food Security Survey Module (HFSSM) consisting of 18 questions (10 questions for households without children). Household food security status is assessed based on a count of the respondent's affirmative answers to those 10-18 questions, standardized into 29 discrete, scalar-valued FSSS, and are generally grouped into three ordinal categories (food security, low food security, and very low food security). The combined population shares in the latter two categories represent the estimated prevalence of food insecurity; the distinction between those two categories provides a measure of the severity of food insecurity.[1]

The annual CPS HFSSM is administered to the CPS respondents up to twice, providing short-term persistence of food insecurity. However, several longitudinal household surveys have implemented (at least part of) the HFSSM tracking longer-term food security status. For example, the nationally representative Panel Study of Income Dynamics (PSID) has implemented HFSSM for seven waves (1999, 2001, 2003, 2015, 2017, 2019, 2021), although there exists a significant gap from 2003-2015. The Early Childhood Longitudinal Study (ECLS-K) - a nationally representative sample of US children – has implemented the HFSSM since 1998, tracking the same children up to five times (1998-2007 for the class of 1998-1999, 2010-2016 for the class of 2010-2011). Some surveys implemented a shorter version by including only a subset of HFSSM questions including the Survey of Income and Program Participation (SIPP) and the Health and Retirement Study (HRS).[2] These data sets have been extensively used to study US food security and its dynamics, such as the intergenerational transmission, persistence, and entry into or exit from food insecurity (Ribar and Hamrick 2003; Hofferth 2004; Kim and Frongillo 2009; Wilde, Nord, and Zager 2010; Ryu and Bartfeld 2012; Kennedy et al. 2013; Gregory et al. 2015; Ziliak and Gundersen 2016; Gundersen et al. 2019; Kim-Mozeleski et al. 2019; Wimer et al. 2019; McDonough et al. 2020; Tiehen et al. 2020; Levy 2022; Insolera 2023; McDonough and Millimet 2024).

While the literature based on the existing FSSS data is rich and informative, it remains limited in scope for two reasons. First, it is difficult to study the persistence of food insecurity. The annual CPS-FSS data yield a maximum of two observations of any survey respondent (beginning in 2001), thus limiting study of household food security persistence to a one-year interval. The other longitudinal surveys that include the HFSSM offer only (i) a limited and discontinuous time series of FSSS observations (PSID), (ii) are not fully nationally representative (ECLS-K), or (iii) use abbreviated HFSSM modules (SIPP, HRS). Second, none can extend back beyond the introduction of HFSSM and FSSS in 1995.

---
[1] FSSS is not only widespread in the United States. It provides the foundation of the Food Insecurity Experience Scale (FIES) that has become one of two food insecurity indicators used to track progress against Sustainable Development Goal number 2: zero hunger (FAO, 2023).
[2] USDA ERS lists surveys that include the HFSSM at https://www.ers.usda.gov/data-products/food-security-in-the-united-states/documentation/.

The first problem has been addressed, in part, by Lee et al. (2024, LBH hereafter) Using the PSID, LBH developed a method to measure food insecurity dynamics by estimating the probability that a household's observed (or imputed) monthly food expenditures equal or exceed the minimal cost of a healthful diet, as reflected by the USDA's Thrifty Food Plan (TFP) - the probability of food security (PFS). LBH, however, constructed household-level measures for the 2001-17 period, using only the balanced panel of households whose head did not change after the PSID moved to biennial frequency in 1999. This paper develops an extension of the LBH method that, using the PSID, extends the measurement of the PFS into a 40-year (1979-2019) panel data series. This method allows us to make three contributions to the literature: we provide: (i) a new source of information on food insecurity dynamics (addressing the first limitation described above); (ii) an estimate of food insecurity, and food insecurity dynamics that precedes 1995 (addressing the second limitation); and (iii) a characterization of the patterns of food insecurity – its prevalence and severity and its demographic correlates over a 40-year period. We estimate food security status in pre-1995 to being higher compared to post-1995 status on average, reflected in higher PFS (>0.8) and lower food insecurity prevalence rates. We also find that the estimated food security status is largely consistent with the official food security status (86%), while the inconsistency is largely driven by those marginally food secure under the official status (i.e., those that affirm some food insecurity questions but not enough to be classified as food insecure). We find that *estimated* food insecurity spells are roughly equally likely to be transitory or persistent. Business cycle effects are strongly associated with both individuals starting *estimated* transitory food insecurity spells and longer duration spells among those already *estimated to be* food insecure. These dynamics were especially pronounced in the aftermath of the Great Recession and explain why estimated food insecurity was higher in the 2010s than it was in earlier decades. The subpopulations likely to persist in *estimated* food insecurity are women, non-White, physically disabled, or less well-educated than the overall population.

2. **PFS motivation and construction**

LBH constructs a new measure of food security, defined as the estimated conditional probability that a household's observed food expenditures equal or exceed the minimal cost of a healthful diet, as reflected by the TFP cost, the PFS. The PFS is a complement to the FSSS - not a substitute -that enables the estimation and study of food security dynamics over extended periods in data sets where measures derived from the FSSS data are unavailable. Unlike the FSSS, which requires a separate survey module, the PFS can be estimated from food expenditures and other household characteristics commonly collected in most household surveys. Thus, the PFS can be used to assess estimated food security status using existing longitudinal survey data even when the FSSS estimates are not available, which necessarily includes all data series before 1995. We emphasize that PFS offers only a prediction of food insecurity. But it tracks the official experiential measure sufficiently well to estimate food insecurity dynamics that cannot presently be studied using the official measure. LBH shows that PFS tracks FSSS much better than does realized food expenditures, which are less strongly correlated with food security status than casual observers might believe (Gundersen and Ribar 2011).

The PFS is estimated in a three-step process to compute the conditional density of household food expenditures for each household and survey period. In the first step, the conditional mean of household per capita food expenditures in the month prior to the survey[3] is regressed on a polynomial of its prior period value - thereby allowing for nonlinear dynamics - and other covariates,

$$W_{ist} = \sum_{\gamma=1}^{2} \pi_\gamma W_{is,t-2}^\gamma + \Lambda X_{ist} + \Omega_s + \omega_t + u_{ist}, \qquad (1)$$

where $W_{ist}$ is the monthly average per capita food expenditures for individual *i* in state *s* reported in year *t*. We construct this dependent variable by dividing the monthly household food expenditure for *i*'s household in year *t* by the number of household members. $X_{ist}$ is a vector of covariates that are understood to be associated with food security, including demographics and socioeconomic status of the reference person (age, sex, race, educational attainment, marital status, employment and disability) and household information (income, family size, ratio of child). We include year and state fixed effects. We do not include individual fixed effects because doing so would yield PFS estimates that vary among individuals within a household in a given year although the underlying PSID and Thrifty Food Plan data used to estimate PFS are household-and-year-specific, thus should not allow for within-household-and-year interpersonal variation.

The predicted value of the outcome variable, $\widehat{W}_{ist}$, is the conditional mean of the individual per capita food expenditure for that period. Given a mean zero error term, $E[u_{ist}] = 0$, the expected value of the squared residuals equals the conditional variance of monthly per capita food expenditures for individual i, state s and year t, $V[W_{ist}] = E[|\hat{u}_{ist}^2|] = |\hat{\sigma}_{ist}^2|)$.

In the second step, the squared residuals from the conditional mean equation are regressed on the same covariates used in equation (1), yielding a regression equation for the conditional variance of per capita food expenditures.

$$\hat{u}_{ist}^2 = \sum_{\gamma=1}^{2} \Pi_\gamma W_{is,t-2}^\gamma + \lambda X_{ist} + \Delta_s + \delta_t + \eta_{ihst} \qquad (2)$$

The third step uses the individual-and-period-specific conditional mean and variance estimates to construct an individual-year-specific conditional cumulative density function (CDF). Assuming $W_{ist} \sim Gamma(\alpha, \beta)$, we calibrate the parameters using the method of moments such that $\left(\alpha = \frac{\widehat{W}_{it}^2}{|\hat{\sigma}_{it}^2|}, \beta = \frac{|\hat{\sigma}_{it}^2|}{\widehat{W}_{it}}\right)$.

The probability of food security (PFS) is 1 minus CDF below minimum threshold food expenditures level, i.e., the estimated probability that $W_{it}$ is no less than $\underline{W_{it}}$, the TFP diet cost specific to that individual's household composition and survey date, as reflected in equation (3) below. This individual-and-year-specific, probabilistic measure, given by equation (3) below, is necessarily a bounded, continuous variable in the [0,1] interval, with a higher probability indicating greater food security. LBH categorizes individuals as food insecure if the estimated probability is below a certain threshold probability. For the period covered by the LBH paper (1995-2019), year-specific thresholds are used to ensure that the share of food insecure individuals in the PSID data matches the individual-level food insecurity prevalence rate as reported by USDA.

---

[3] PSID permits respondents to choose the recall period over which they report food expenditures. The overwhelming majority choose weekly recall, with monthly recall the next most common. Given the month-specific TFP cost against, which PFS compares the estimated food expenditures distribution (see below), we convert all food expenditures to monthly flows and match that with the TFP corresponding to the survey month and household demographic composition. Each household has one such observation per year, which we assign to any sample individual resident in that household that year.

$$PFS_{it} = Pr\left(W_{it} \geq \underline{W_{it}}|\Theta\right) = 1 - F_{W_{it}}\left(\underline{W_{it}}|\Theta\right) \in [0,1] \qquad (3)$$

Applying the LBH method to cover the full 40-year period requires several adjustments. First, LBH estimated PFS at the household-year-level, using a balanced panel of households where the identity of the head remained unchanged over time. Extending the panel back in time requires allowing the household head - or 'reference persons" (RP) in PSID terminology – to change and thus requires a method of linking RPs over time. Otherwise, the sample becomes increasingly non-representative, undercutting a key reason to use the nationally representative PSID. Section 3 explains in detail how we address this. Second, while the LBH used a generalized linear model (GLM) logit link regression under Gamma distributional assumption, this paper uses Poisson quasi-MLE, which is consistent for any non-negative response variables (Wooldridge 1999). Third,[4] LBH defined year-specific cut-offs such that the share of food insecure individuals in their data matched the prevalence of food insecurity as measured by the FSSS. Like LBH, we set $\underline{W_{it}}$ to exactly match the official food insecurity prevalence rate (based on CPS-FSS) for the period 1995-2019. But because FSSS-based food insecurity prevalence estimates do not exist prior to 1995, for the period 1979-1994, we estimate a simple regression model relating year-specific PFS cut-offs, 1995-2019, to a suite of macroeconomic variables, and then use that regression model to predict cut-offs in the pre-1995 period.

3. **Panel data series construction and sample descriptive statistics**

Starting with 18,000 individuals from 4,800 households in 1968, the PSID surveyed 82,000 individuals from about 9,000 households over 41 waves as of 2019, annually until 1997 and biennially since then. Since its initial survey in 1968, the PSID has followed the household heads – called reference persons (RP) since 2017 – surveyed in 1968 as well as those who are genealogically related to them (i.e., their children, grandchildren, etc.). The PSID collects individual-level information (e.g., household role, demographics, socioeconomic status) as well as information (e.g., food expenditures, Supplemental Nutrition Assistance Program (SNAP) participation) on the household in which the individual resided at the time of interview.[5] Researchers have used this 50-year long food expenditures data series to study various topics such as welfare programs (Knaub 1981; Senauer and Young 1986; Hoynes and Schanzenbach 2009; Kim and Shaefer 2015), grocery taxes (Wang et al. 2023), labor market participation (Shotick 2014), and heterogeneity in food expenditures (Gupta et al. 2021). The PSID has also collected data used to estimate the FSSS over seven survey rounds (1999, 2001, 2003, 2015, 2017, 2019, 2021), allowing researchers to study food security over periods as well as across generations (Hofferth 2004; Kim-Mozeleski et al. 2019; Corman 2022; Insolera et al. 2022; Lee et al. 2024).

The appeal of using PSID to construct a time series estimate of food security is twofold. First, it is the longest nationally representative household panel data set in existence. Second, Tiehen et al. (2020) validated the PSID data against CPS-FSS, endorsing its use for the study of food security patterns nationwide. But in order to maintain the nationally representativeness of the sample, one must be careful in constructing the panel. Researchers using PSID to investigate household-level status over time create a household-level panel by defining a household over time based on specific inclusion criteria, such as limited to no changes in the entire household roster, no change in the RP, or no change in the RP and spouse. For example, LBH used the balanced panel of roughly 2,700 households whose RP did not change over the 1999-2017 period since PSID converted to biennial frequency.

---

[4] Data and codes used to construct the data are available at [URL to be added after acceptance].
[5] Strictly speaking, PSID collects information on a "family", which differs from a "household," which in PSID is a location-based definition that can include more than one family residing in a single housing unit. However, as of the 2021 PSID survey wave, more than 92% of households consist of a single family. Therefore, we use the term "household" synonymously with "family," as is common in the literature.

But that approach can suffer from significant attrition bias the longer one extends the panel. Of the 6,007 RPs surveyed in 1977 – the first year in our study since we use lagged food expenditures to construct PFS – only 667 (11%) remained the RP in the same family through 2019 and those RPs were all late 50s or older by 2019, yielding a subsample that dramatically underrepresents younger Americans. So here we follow a different approach, tracking children who split off from their parents to form their own families in adulthood.

More specifically, we construct a panel data set of approximately 270,025 observations by tracking 17,862 individuals over 26 survey waves (1979 - 2019). We assign household-level variables from PSID, such as food expenditures or SNAP receipt, to individuals within those households and link the household-level observations over time as spouses, children or grandchildren co-resident in original PSID households move across households over time. Individuals are therefore associated with the characteristics of the (potentially different) household(s) in which they reside in each survey round. We include all individuals who were household members in the original 1968 PSID panel and were surveyed in 1977, as well as those genealogically related to them, including children born/adopted to the initially surveyed, but not stepchildren because the PSID does not follow those who are not genealogically related, giving such individuals zero survey weight. We further narrow the sample of included individuals to those who were either the RP or a spouse in the household units in which they resided at least once from 1977 to 2019. Other household members' household-level outcomes are captured by the RP or RP's spouse, thus there is no loss of information by dropping individuals who were never a RP or spouse. We also restrict analysis to individuals living in one of the 48 continental states or the District of Columbia, for which we have the full Thrifty Food Plan (TFP) cost data over the study period (we do not for Alaska, Hawaii, nor for other US territories outside the contiguous 48 states). Fourth, we do not include the 1990-1992 Latino supplemental and the 2017-19 immigrant refresher samples because they were surveyed for a relatively short period of time. Last, we do not include 1988 to 1991 (four waves) for which we could not construct the outcome variables due to the absence of food expenditures in PSID. We adjust survey weights to capture multiple individuals within the same household by dividing the survey weight by the number of individuals included in the sample within that household. Our sample's household characteristics are similar in composition and show similar trends over the study period, implying that the sample maintains reasonable national representativeness. See Appendix A for more detail and a heuristic example of which individuals are tracked over time in the sample.

Table 1 shows individual-level and individual-year-level summary statistics for the study sample. Variable construction details can be found in Appendix A. 51% of observations come from individuals surveyed in 1977; 49% of the observations are of individuals - children or spouses - who first appeared in later waves. 39% were estimated to be food insecure at least once during the survey period, and the same share of individuals used SNAP at least once over the period. Individual-year-level statistics show that households spent $310 per capita monthly for overall food spending (w/o SNAP benefit). While 7% received SNAP benefits amounting to $99 on average, 18% had a Normalized Monetary Expenditure (NME) – the ratio of food expenditures to the TFP cost for their household composition – less than one, meaning they spent less on food than the TFP cost (Yang et al. 2019). The average Probability of Food Security (PFS, Lee et al. 2024) estimate, the main outcome of our study is 0.81, with 12% below cut-off probability, representing a PFS-based estimate of the 40-year average food insecurity rate nationwide.

**Table 1: Summary statistics**

| Variable | Mean (SD) / (Percent) |
|---|---|
| *(a) Individual level (N=17,862)* | |
| Female (RP) | (53.2%) |
| Surveyed in 1979 | (51.7%) |
| Number of waves surveyed | 15.88 |
| | (7.15) |
| Ever estimated to be food insecure | (38.7%) |
| Ever used SNAP benefit | (26.5%) |
| Years SNAP benefits used | 1.11 |
| | (2.68) |
| *(b) Individual-year level (N=270,025)* | |
| Age (RP) | 48.06 |
| | (16.51) |
| Female (RP) | (21.9%) |
| non-White (RP) | (16.7%) |
| Married (RP) | (68.7%) |
| Education (RP) | |
|   Less than HS | (17.5%) |
|   High School/GED | (35.2%) |
|   Some college | (18.6%) |
|   College | (28.7%) |
| Employed (RP) | (71.6%) |
| Disabled (RP) | (19.0%) |
| Family size | 2.94 |
| | (1.54) |
| Proportion of children | 0.22 |
| | (0.26) |
| Region | |
|   Northeast | (10.7%) |
|   Mid-Atlantic | (14.5%) |
|   South | (26.5%) |
|   Midwest | (26.2%) |
|   West | (22.2%) |
| Received SNAP | (7.1%) |
| Annual family income per capita (K) (Jan 2019 dollars) | 33.70 |
| | (28.79) |
| Monthly food expenditure per capita (Jan 2019 dollars) | 310.06 |
| | (189.54) |
| SNAP benefit amount (Jan 2019 dollars) | 98.82 |
| | (59.06) |
| Estimated to be food insecure by PFS | (11.8%) |
| NME < 1 | (18.1%) |

## 4. PFS patterns in the data

Figure 1 shows that across all years and individuals, the mean probability of food security is 0.81, meaning that the average American each year has an 81 percent likelihood of spending more on food than the Thrifty Food Plan budget cost appropriate for their household.[6] Figure 1 shows the annual estimates of mean PFS as well as the bottom 5th and 20th percentiles of its distribution. Individuals in the bottom 5% percentile are unlikely to have food expenditures (probability less than half) equal to or greater than the cost of the TFP. Over the 40 years for which we have data, mean PFS remains within a tightly defined range, from 0.77 to 0.85. It falls during the early 1980s and early 1990s recessions, and especially during the Great Recession. Since that time, the PFS distribution has risen, reflecting improved estimated food security over the decade of post-Great Recession recovery. These PFS estimates are correlated with household characteristics in the ways one would expect; positively associated with family income, employment, and educational attainment, negatively associated with female reference person, family size or disabled reference person.[7]

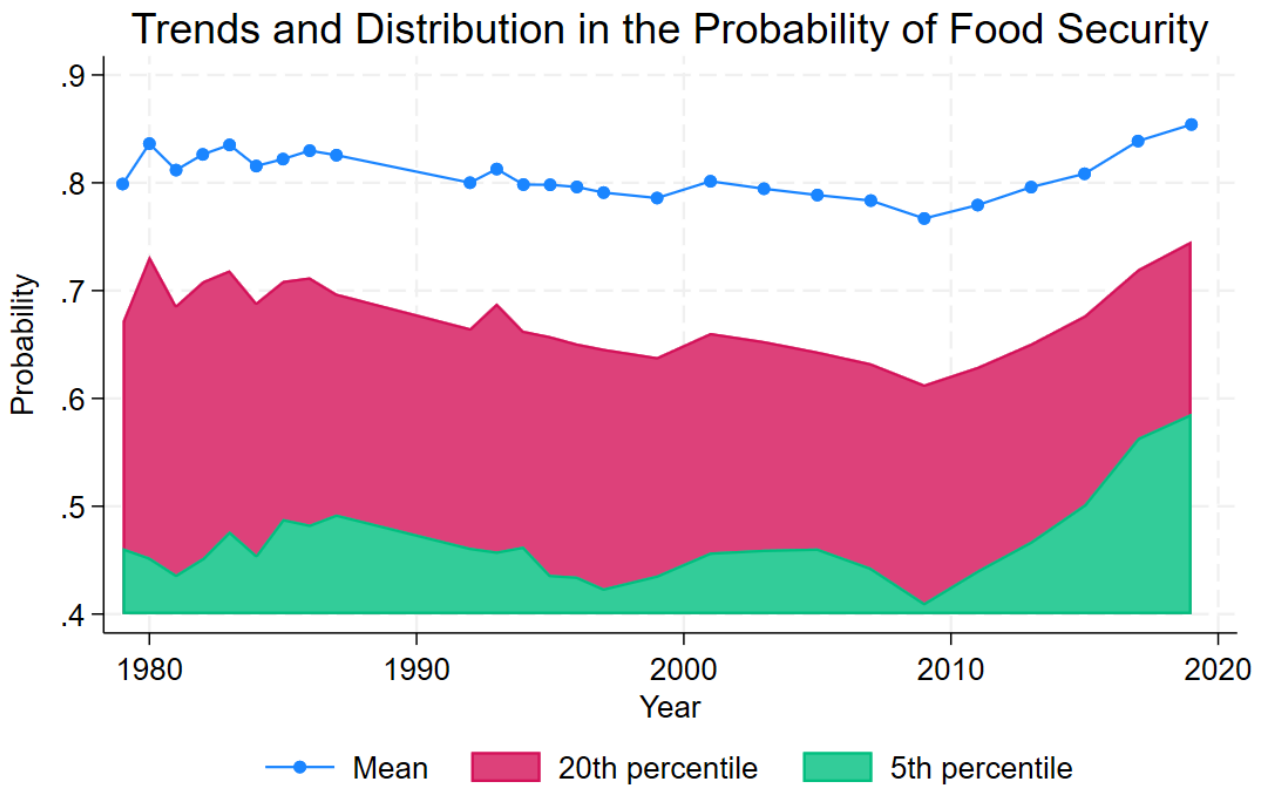

**Figure 1:** The mean (dot), 20th and 5th percentiles of the year-specific PFS distribution for each PSID wave year.

---

[6] Table B1 reports the estimates of equations (1) and (2) - the conditional mean and conditional variance equations - when we apply this method to the 1979-2019 unbalanced panel data.

[7] Full regression results are reported in Table B2.

0

Next, we apply the thresholds ($W_{it}$ in equation (3)) described in section 2, to determine food insecurity prevalence rates over study period. For the period when the official US food security prevalence rates are available (1995-2019), we set the thresholds to exactly match the official food insecurity prevalence rate (based on CPS FSSS), as in LBH. For the 1979-95 period for which no official food security prevalence rates exist, we estimate a simple regression model to relate year-specific PFS thresholds from 1995 to 2019 to national SNAP participation rate[8] and then use that parameterized regression model to generate estimates of food security prevalence. We use the SNAP participation rate because (i) SNAP participation is highly correlated with food security (among many studies, see, for example, Gundersen et al. 2009, Ratcliffe et al. 2011, Gregory et al. 2015, Gordon et al. 2018 ) (ii) moderately explains variations in realized PFS thresholds (see below) and (iii) varies with the business cycle (see Figure B1). We emphasize that the purpose of this regression model is to predict PFS thresholds; it is not inference with respect to any of the explanatory variables, given few degrees of freedom. In addition to this regression-based thresholds estimation, we further provide two boundary thresholds which estimate households whose PFS is in the bottom 5% and 20% as food insecure each year. These bounds are based on historical U.S. individual food insecurity rates that fluctuated between 10 to 20%, thus our bounded estimates serve as possible maximum/minimum estimated food security prevalence over the study period.

Table 2 shows the regression results for predicting the PFS threshold. Columns (1)-(4) report bivariate relationships between the PFS threshold and each indicator we tested (Hake et al. 2024).[9] Column (1) shows that per capita disposable income is positively associated with higher food expenditures, increasing overall PFS thus is strongly associated with the threshold PFS value to be classified as food secure ($R^2$=0.67). SNAP participation (column 2) is positively associated with PFS thresholds – and has moderate explanatory power; ($R^2$=0.29) – because SNAP participation is associated with greater (SNAP-inclusive) food expenditure, increasing overall PFS, and moves inversely with the business cycle, rising during recessions and falling during periods of robust economic growth. Despite their strong association with the business cycle, the unemployment rate (column 3) and GDP per capita growth rate (column 4) exhibit non-significant association with the PFS threshold. Column (5) reports the multivariate regression of the PFS threshold on all four indicators. This model has the highest $R^2$, but with considerable potential risk of overfitting (4 covariates with just 14 observations).

The blue line in Figure 2 depicts the realized PFS thresholds anchored to the official USDA food security prevalence rate, while the other five lines in the figure show the predicted thresholds from columns (1)-(5) in Table 2. The grey (income only) and black line (full model) are mostly monotonically increasing over the period, yielding very low PFS thresholds in early years, with the mechanical implication of very low food insecurity prevalence early in the study period, which seems unlikely. These patterns are driven by the relatively high explanatory power of real per capita income, which increases steadily over the period. Despite that variable's relatively high power in predicting PFS thresholds, income is therefore not a good indicator to predict PFS thresholds. The other three predictions – based on SNAP participation rates, unemployment rates and per capita GDP growth – display similar trends in the pre-1995 period, but the one based on SNAP participation both fits the data best and reflects business cycle patterns during recessions, so we favor using it to estimate 1979-1994 PFS thresholds for classifying households as food secure or insecure.

---

[8] National SNAP participation rate is imputed through dividing SNAP participating population by estimated total U.S. population.
[9] Table B3 reports the correlation coefficient matrix among full set of macroeconomic variables tested.

Table 2: PFS Thresholds and Macroeconomic Indicators, 1995-2019

|  | 1 | 2 | 3 | 4 | 5 |
|---|---|---|---|---|---|
| ln(disposable personal income per capita - 2017 dollars) | 0.259*** (0.058) |  |  |  | 0.172** (0.065) |
| SNAP Participation Rate (%) |  | 0.008** (0.003) |  |  | 0.006* (0.003) |
| Unemployment Rate (%) |  |  | -0.002 (0.006) |  | -0.013* (0.006) |
| Annual GDP per capita growth rate (percent) |  |  |  | -0.004 (0.006) | -0.007 (0.005) |
| Intercept | -2.144*** (0.613) | 0.508*** (0.029) | 0.603*** (0.045) | 0.595*** (0.012) | -1.210 (0.677) |
| Number of observations | 14 | 14 | 14 | 14 | 14 |
| R-squared | 0.67 | 0.29 | 0.01 | 0.03 | 0.77 |

*** p<.01, ** p<.05, * p<.1

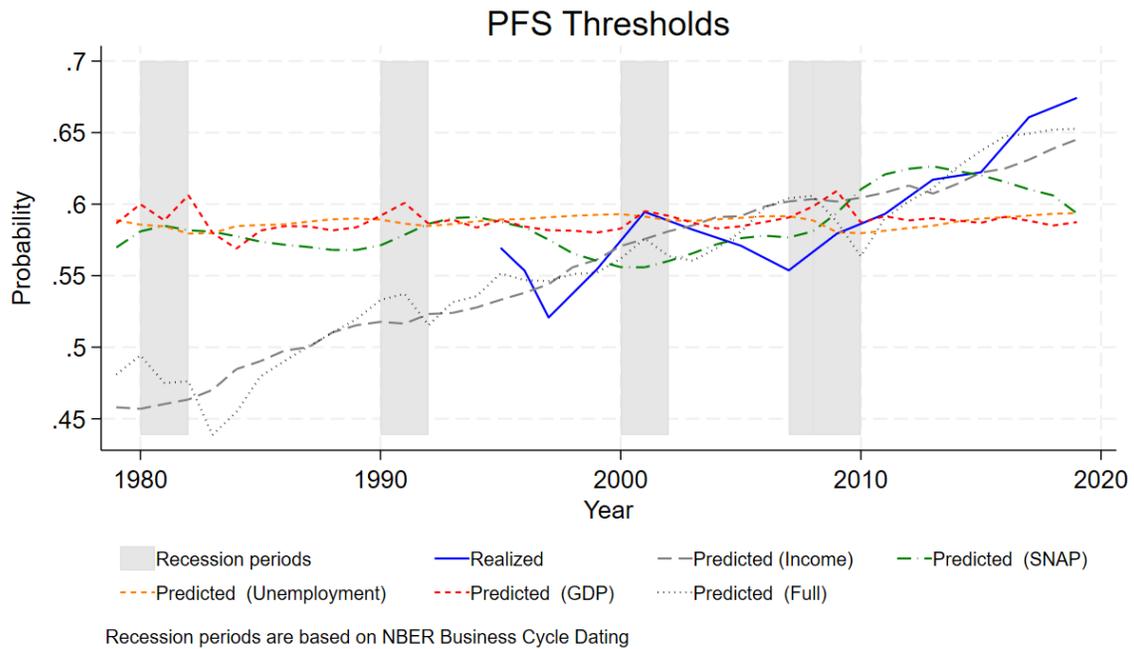

**Figure 2**: PFS Food Security Thresholds, 1979-2019

Across the forty years of data, the average prevalence of estimated individual food insecurity is 11%. Figure 3 shows estimated food insecurity prevalence rates resulting from the PFS estimates and estimated PFS food security thresholds (Table 2 and Figure 2). Over the period 1979-2019, the estimated food insecurity prevalence ranged from a low of 8% percent in 1985 to 17% in 2009. Between 1979 and 1999, estimated food insecurity prevalence varied little, from 8% to 12%. Periods of economic expansion, most notably during the 1990s, are associated with a reduction in the prevalence of food insecurity as estimated by the PFS but these are never sufficient to cause the estimated prevalence to fall below 10%. The Great Recession brought a sharp increase in the estimated prevalence of food insecurity – the 16% figure for 2009 is the highest estimated prevalence over the 40-year period covered by our data - but unlike previous recessions, the return to pre-recession food insecurity prevalence was slower (Leete and Bania 2010). We return to this below.

Figure 4 disaggregates estimated PFS by sex, race, and educational attainment.[10] These box-and-whisker plots show the interquartile (25th-75th percentile) range of period-average individual estimates in the box, with the median value indicated by the horizontal line within the box, and maximum and minimum values in the whisker ends. There are four groups, ordered from left to right by the individual's ultimate educational attainment, from less than a high school diploma on the left, to a college degree on the right. Within each block we display pairs of distributions of PFS estimates for White individuals on the left and non-White individuals on the right, with males in blue and females in red.

Consistent with previous studies (Carlson et al. 1999; Broussard 2019; Heflin et al. 2022; Lee et al. 2024; Rabbitt et al. 2024), the data reveal striking disparities in estimated food security status. PFS is strongly positively correlated with educational attainment. The lowest median PFS is found among individuals who have not completed high school and the highest is found among individuals who have completed college. Within each educational category, median PFS is higher for men compared to women and is higher for Whites compared to Non-Whites; these sex (racial) differences widen(narrow) as educational attainment increases. Non-White women with less than a high school education have a median PFS of 0.55, while White men with a college education have a median PFS of 0.95. The large gap between these two groups illustrates the fact that the estimated food (in)security experience of Americans differs dramatically based on intersecting educational attainment, sex and racial characteristics.

---

[10] Appendix A3 describes how we treated missing values in individual race and educational attainment.

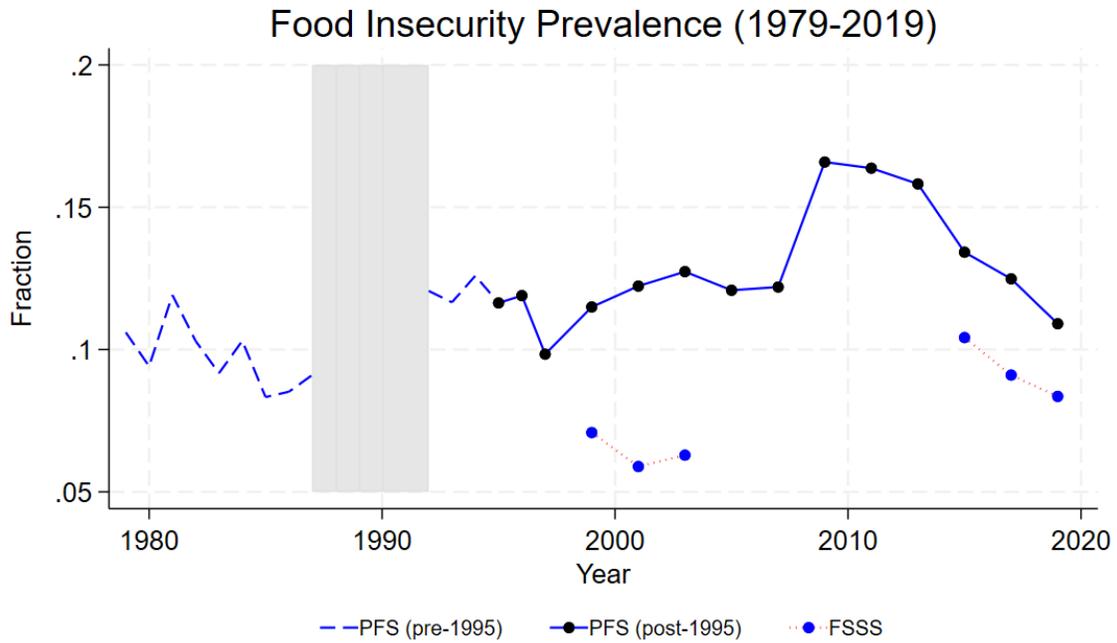

Post-1995 prevalence is anchored to the official USDA individual prevalence.
PFS is missing from 1988 to 1991 due to missing data in PSID.

**Figure 3**: Estimated Food Insecurity Prevalence, 1979-2019

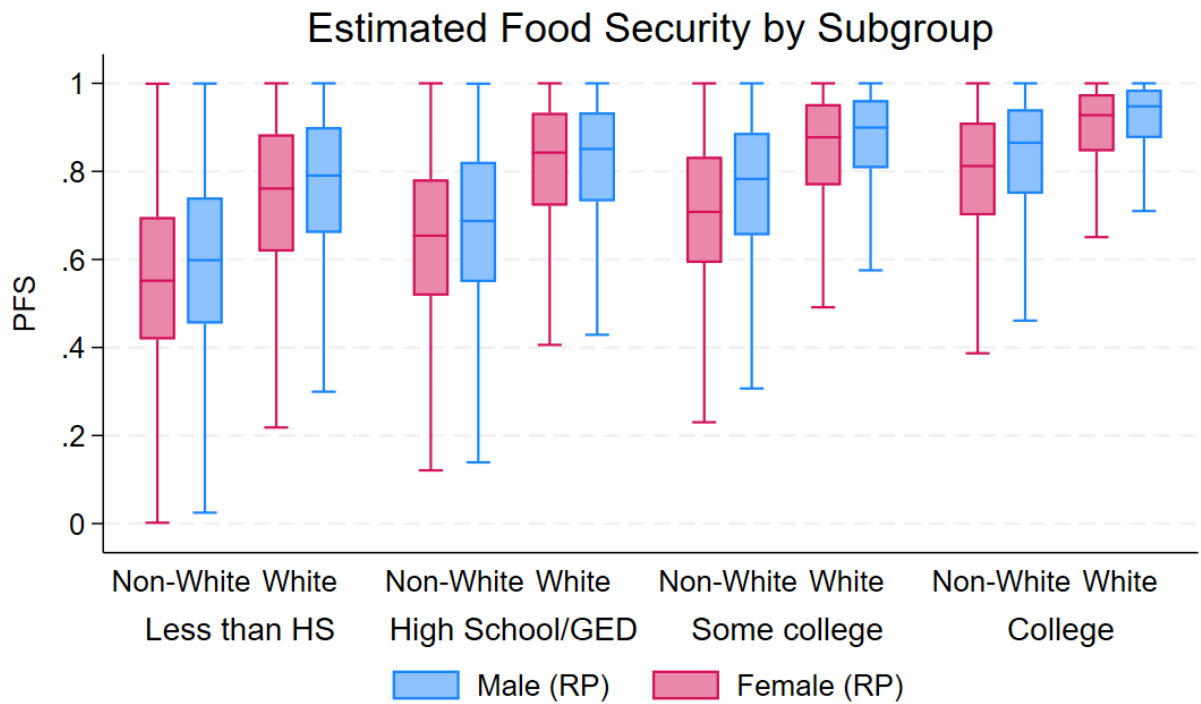

**Figure 4:** Distribution of 1979-2019 average PFS estimates based on individual educational attainment, sex, and race.

We note that the FSSS-based food insecurity rate (for the years for which this is available in the PSID) is lower than PFS-based estimates. This lower rate is consistent with Tiehen et al. (2020) who report that the PSID food insecurity rate is lower than that in the CPS-FSS, on which the USDA official rate is based. Because we anchor PFS to the official food insecurity prevalence estimates that USDA-ERS has generated each year since 1995 from the CPS-FSS, these PFS estimates track the official national prevalence estimates, while the FSSS estimates generated from PSID do not. The PFS measure nonetheless tracks the FSSS-based measure reasonably well in the shorter sample for which both can be calculated. The estimated rank correlation coefficients between PFS and FSSS over the six years we have for both in PSID are 0.36 (Spearman) and 0.28 (Kendall's tau-b), significantly different from zero. Table 3 shows how the two measures categorize individuals as food insecure (or not) each year. Overall, 86% of individuals are categorized the same (food secure or food insecure) under both PFSS and FSSS, with match rates varying only in the 84-88% range. On the other hand, among the 8% of individuals food insecure under the FSSS, only 37.5% are estimated to be food insecure under the PFS (3%).

**Table 3: Estimated Food Security Status as estimated by PFS and FSSS**

| Status measured by PFS / | 1999 | 2001 | 2003 | 2015 | 2017 | 2019 | Total |
| --- | --- | --- | --- | --- | --- | --- | --- |

| Status measured my FSSS | | | | | | | |
|---|---|---|---|---|---|---|---|
| Food secure / Food secure | 0.84 | 0.85 | 0.84 | 0.80 | 0.82 | 0.83 | 0.83 |
| Food insecure / Food insecure | 0.03 | 0.03 | 0.03 | 0.04 | 0.03 | 0.03 | 0.03 |
| Food insecure / Food secure | 0.09 | 0.09 | 0.10 | 0.10 | 0.09 | 0.08 | 0.09 |
| Food secure / Food insecure | 0.04 | 0.03 | 0.03 | 0.07 | 0.06 | 0.06 | 0.05 |

We use two approaches to investigate the 14% of mismatches. First, for those classified as food insecure using PFS but food secure by FSSS, we re-classified their status using the ordered FSSS. That is, if the USDA food insecurity prevalence rate is 11.5% for a given year, we classify the 11.5% highest FSSS scores as food insecure to mechanically match the CPS-FSS-based prevalence rate, as in LBH. This resolves the problem of PSID FSSS data yielding food insecurity prevalence estimates well below the official rate. In this approach, nearly 2,800 food secure households (5% of sample) under the PFS were re-classified as food insecure. The overall matching rates were very similar, as reported in Table B4.

Second, we compare demographic and socioeconomic status with those classified under different categories and those under the same categories. Table 4, panel (a) shows the share of different classifications by characteristics. We found that those who are known to be more vulnerable to food insecurity – female, non-White, disabled, and low educational attainment – are more likely to be differently classified by PFS and FSSS. For instance, 33% of those without a high school diploma are differently classified by PFS and FSSS.

To further examine what largely drives these mismatches, we examined summary statistics for each classified group, as shown in panel (b). We hypothesize that those who are differently classified by FSSS only - (1) and (2), (3) and (4) - should have similar food expenditure levels as PFS is an expenditure-based measure, while other characteristics associated with food security (such as vulnerability to future income shock) may differ. Similarly, those differently classified by PFS only – (1) and (3), (2) and (4) – should be largely different in food expenditure, but not so different in other characteristics. We found that the difference between (1) and (2), and (3) and (4) are not so significant in food expenditure, but very large in terms of vulnerable characteristics – sex, age, race, disability and low educational attainment. These differences are consistent with our first hypothesis. Our second hypothesis is partially satisfied; the difference between (2) and (4) in characteristics are not so large except food expenditure as hypothesized, but the differences are large between (1) and (3). These large differences are driven by those who are marginally food insecure under the FSSS (raw score=1 or 2). 71% of them are estimated to be food insecure under the PFS, and they account for only 6% in (1) but 22% in (3). Their characteristics are very different from those who are in (1), but closer to those in (4).

We summarize our findings from Table 4 as below. First, PFS is more capable of estimating food security status on average and for those more likely to be food secure, compared to estimating those less likely to be food secure. Second, the mismatches between the PFS-based status and FSSS-based status are largely driven by those with adequate recent food expenditures but who are more vulnerable to shocks to future income due to demographic characteristics and income. Such individuals may be more likely to worry about future food consumption and thus are perhaps more likely to affirm the questions in the food security survey module, leading to their classification as food insecure under the FSSS despite their adequate current food expenditures. This finding underscores a key difference between the PFS measure based on objective food expenditure data and the FSSS measure based on subjective assessment of a suite of conditions. There will be a subpopulation for whom the two measures do not fully correspond. It is unclear which measure one should favor.

### Table 4: Summary Statics by PFS/FSSS Status

|  | (1) Food secure (PFS and FSSS) | (2) Food secure (PFS) / Food insecure (FSSS) | (3) Food insecure (PFS) / Food secure (FSSS) | (4) Food insecure (PFS and FSSS) |
|---:|---:|---:|---:|---:|
| Total | 0.83 | 0.05 | 0.09 | 0.03 |
| Male (RP) | 0.88 | 0.04 | 0.06 | 0.02 |
| Female (RP) | 0.68 | 0.08 | 0.18 | 0.07 |
| White (RP) | 0.88 | 0.04 | 0.06 | 0.02 |
| non-White (RP) | 0.61 | 0.07 | 0.24 | 0.09 |
| Not married (RP) | 0.89 | 0.03 | 0.06 | 0.02 |
| Married (RP) | 0.73 | 0.08 | 0.14 | 0.05 |
| Not disabled (RP) | 0.86 | 0.04 | 0.08 | 0.02 |
| Disabled (RP) | 0.72 | 0.08 | 0.13 | 0.06 |
| Less than HS (RP) | 0.56 | 0.07 | 0.26 | 0.12 |
| College (RP) | 0.95 | 0.02 | 0.02 | 0.01 |
| (a) Classification status by RP characteristics | | | | |
| N | 37,839 | 2,992 | 8,555 | 3,132 |
| Female (RP) | (19.3%) | (37.6%) | (46.9%) | (49.6%) |
| Age (RP) | 51.50 | 42.95 | 48.57 | 41.83 |
|  | (16.56) | (14.78) | (19.72) | (14.28) |
| non-White (RP) | (14.2%) | (27.0%) | (50.0%) | (55.2%) |
| Married (RP) | (68.9%) | (39.8%) | (45.6%) | (41.0%) |
| Disabled (RP) | (16.5%) | (30.7%) | (28.0%) | (39.0%) |
| Less than HS | (7.3%) | (15.1%) | (31.4%) | (40.6%) |
| Family size | 2.58 | 2.34 | 3.62 | 3.69 |
|  | (1.31) | (1.40) | (2.01) | (2.21) |
| ln(per capita income) | 10.46 | 9.79 | 8.94 | 8.68 |

|  | (0.71) | (0.64) | (0.97) | (0.92) |
| --- | --- | --- | --- | --- |
| Food expenditure per capita (including SNAP benefit) | 359.00 | 304.78 | 192.24 | 197.13 |
|  | (207.23) | (196.17) | (125.98) | (141.11) |
| PFS | 0.86 | 0.77 | 0.51 | 0.47 |
|  | (0.11) | (0.10) | (0.10) | (0.13) |
| FSSS (raw score) | 0.08 | 5.62 | 0.32 | 6.04 |
|  | (0.34) | (2.43) | (0.64) | (2.99) |

(b) Summary statistics for each categorized group

## 5. Estimated Food Security Dynamics

LBH describes two different approaches to constructing estimated food security dynamics measures from PFS estimates. Here we use the spells approach, which reflects the number of consecutive survey periods in which an individual's PFS falls beneath W, the normative standard level that roughly equates the prevalence under PFS over time with the official prevalence rate based on the FSSS. The other, permanent approach (which we do not discuss here due to space constraints) characterizes period-average experiences over a time interval, enabling identification of estimated transitory food insecurity experienced by individuals who are on average estimated to be food secure over the time interval, as well as transitory moments of estimated food security experienced by the estimated chronically food insecure, defined as those who are on average estimated to be food insecure over the time interval.

Our PFS estimates generate three key descriptive findings. First, the average spell length of estimated food insecurity is 3.13 survey waves (recall that a spell corresponds to a one-year period when the PSID was surveyed annually (pre-1997), and a 2-year period when the PSID was done biennially (post-1997) with relatively little differences in mean spell length by, education, race, or location. Second, the average spell length disguises considerable heterogeneity in spell lengths as shown in Figure 5. Among the 38.7% of our study sample who experienced at least one food insecurity spell, approximately 52% of the estimated food insecurity spells last a single survey wave and another 17 percent last two survey waves suggesting that about 69% of spells are transitory (2 years or less) and 31% are more persistent (3 years or more).[11]

Further, short-term estimated food security dynamics – estimated food security status in two consecutive rounds - differs by person type (Table 5). Table 5 is constructed by aggregating all transition matrices in estimated food security status across two waves[12] (estimated to be food secure in both waves, estimated to be food insecure in the first wave (but estimated to be food secure in the second wave), estimated to be food insecure in the second wave (but food secure in the first wave), and estimated to be food insecure in both waves) between 1979 and 2019. Fully 84% of these respondents were estimated to be food secure in both waves, 4% were estimated to be food insecure in the first wave only, 4% were estimated to be food insecure in the second wave and 8% were estimated to be food insecure in both waves. Fewer individuals were food secure in two consecutive waves in later periods. Consistent with our finding that the PFS is lower for women, non-Whites, those with less than a high school education and those who are disabled were all more likely to be estimated to be food insecure in both waves of these transition matrices. The fact that non-White individuals have higher rates of both "Insecure in 1st round only" and "Insecure in 2nd round only" than White individuals is consistent with McDonough and Millimet (2024), who found that non-White households have greater mobility, both upward and downward.

---

[11] The share estimated to be food insecure in all 26 survey waves is higher than those with slightly shorter spell lengths because longer spell lengths – those that began prior to 1979 (left-censoring) or that extended after 2019 (right-censoring) – are necessarily truncated at 26 survey rounds. Censoring at the upper bound is a weakness of the spell-length approach (Lee et al. 2024).

[12] Meaning a 2-year period for pre-1997, and 4-year period for post-1997.

Figure 6 visualizes these dynamics over time by disaggregating, by wave, the percentage of individuals who were both food insecure (FI) in the current and prior wave - "Still FI" – and those who become food insecure, having been food secure in the prior wave – "Newly FI". (Note that in every wave, there are a small number of individuals whose prior food security status is unknown.) The percentage of "Still FI" individuals remains remarkably constant between 1981 and 2007. Over this period, changes in estimated food insecurity are driven largely by individuals who become newly food insecure. The Great Recession (2007-2009) causes the largest increase in newly FI individuals over the 40-year period covered by our data. Once food insecure, it took time for individuals to return to food security as evidenced by the higher percentage of Still FI individuals (relative to the pre-2009 period) in 2011, 2013, 2015, and 2017.

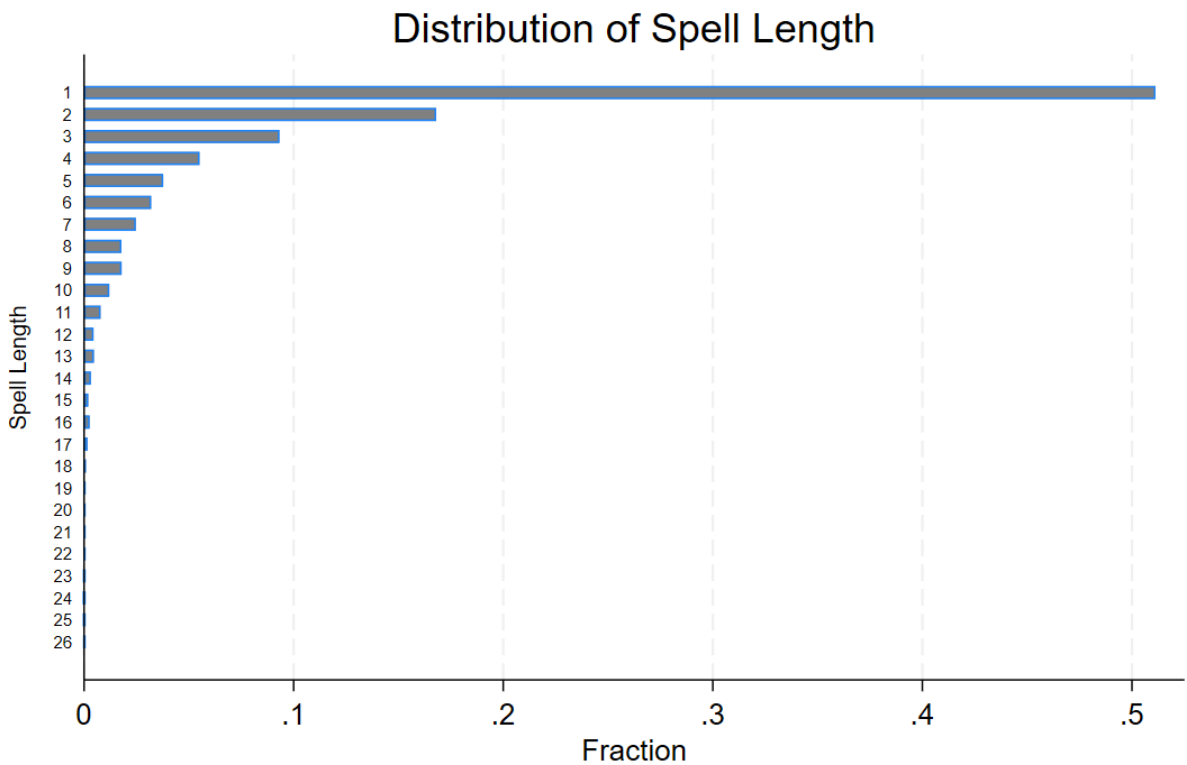

**Figure 5: Spell Length of Estimated Food Insecurity**

Note: The maximum observable spell length in the data is 26, as we described in Section 3, so that observation necessarily includes any spells that were (unobservably) longer.

**Table 5 Transition in *Estimated* Food Security Status, by period, sex, race and education**

| Category | | Number of observations | Insecure in both rounds | Insecure in 1st round only | Insecure in 2nd round only | Secure in both rounds |
|---|---|---|---|---|---|---|
| Total | | 209,095 | 0.08 | 0.04 | 0.04 | 0.84 |
| Period | 1981-1990 | 77,905 | 0.06 | 0.04 | 0.03 | 0.87 |
| | 1991-2000 | 48,411 | 0.08 | 0.04 | 0.03 | 0.85 |
| | 2001-2010 | 43,804 | 0.08 | 0.04 | 0.05 | 0.83 |
| | 2001-2019 | 38,975 | 0.09 | 0.06 | 0.05 | 0.81 |
| Sex | Female | 117,087 | 0.09 | 0.05 | 0.05 | 0.82 |
| | Male | 92,008 | 0.06 | 0.04 | 0.03 | 0.87 |
| Race | Non-White | 80,617 | 0.25 | 0.09 | 0.08 | 0.58 |
| | White | 125,464 | 0.04 | 0.03 | 0.03 | 0.89 |
| Education | Less than High School | 51,193 | 0.19 | 0.07 | 0.07 | 0.67 |
| | High school | 78,859 | 0.08 | 0.05 | 0.05 | 0.82 |
| | College, no degree | 44,779 | 0.05 | 0.04 | 0.03 | 0.88 |
| | College degree | 39,592 | 0.02 | 0.01 | 0.01 | 0.96 |

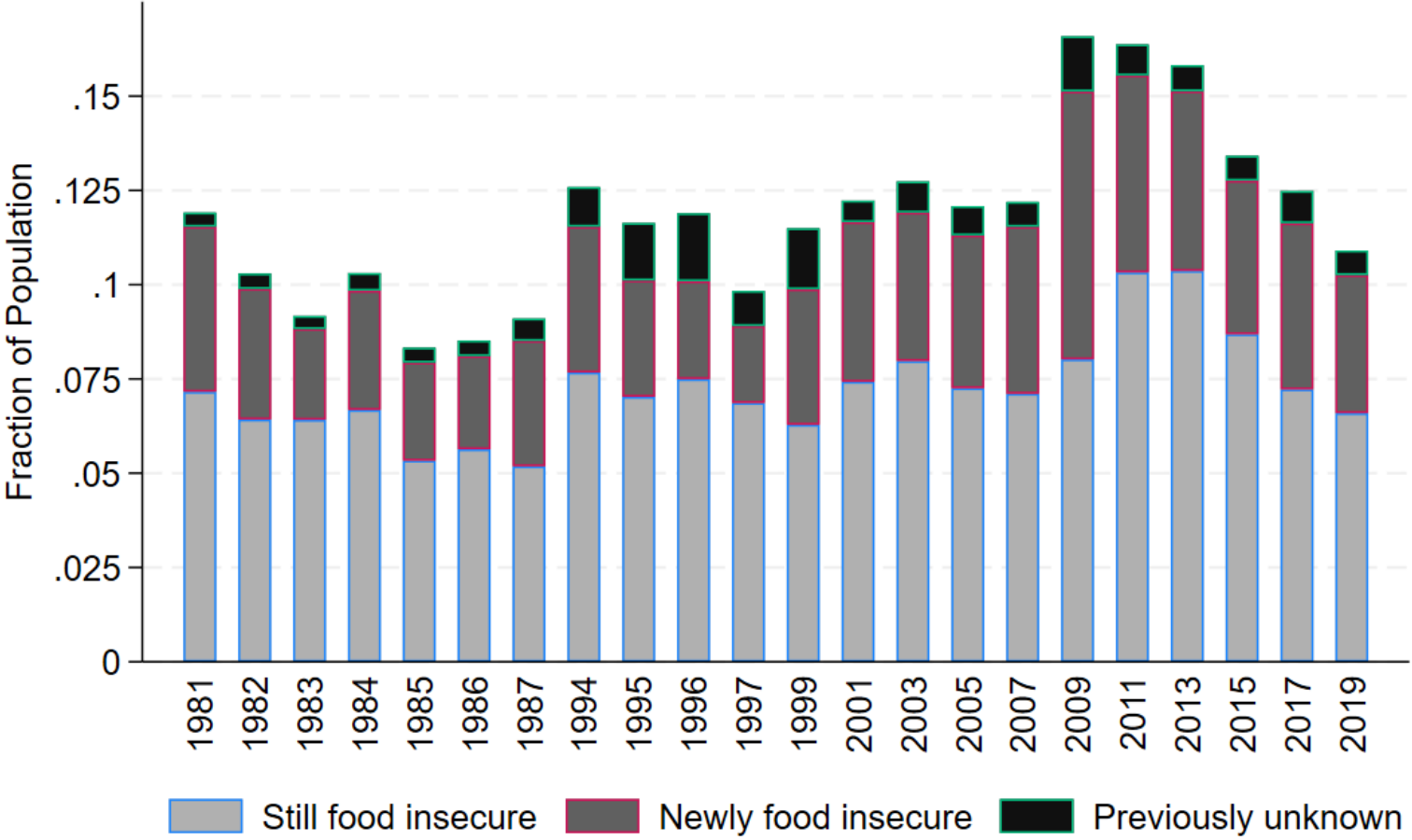

Figure 6: Estimated Food Security Status by type

Table 6 shows the estimated prevalence in chronic food insecurity across different subgroups over decades. Consistent with what we see in Figure 6, across all disaggregation, the prevalence of chronic food insecurity increases in the aftermath of the Great Recession, though the extent of this increase varies by demographic group. Perhaps the most striking is the increase in the likelihood that individuals without a high school diploma, from 0.10 to 0.19 prior to 2010, to 0.26 from 2011 onwards. This is consistent with the fact that: (a) the Great Recession produced a large increase in unemployment for this group; (b) the rate of unemployment for individuals without a high school diploma only returned to pre-Great Recession levels in 2016; and (c) the slow reduction in the prevalence of estimated food insecurity after 2009.

**Table 6: Chronic Food Insecurity - food insecure in at least two consecutive waves**

| Category | | 1981-1990 | 1991-2000 | 2001-2010 | 2011-2019 |
|---|---|---|---|---|---|
| Sex | Female | 0.07 | 0.09 | 0.09 | 0.11 |
| | Male | 0.05 | 0.06 | 0.07 | 0.07 |
| Race | Non-White | 0.23 | 0.27 | 0.24 | 0.24 |
| | White | 0.04 | 0.04 | 0.04 | 0.05 |
| Education | Less than High School | 0.14 | 0.19 | 0.22 | 0.30 |
| | High school | 0.05 | 0.08 | 0.09 | 0.13 |
| | College, no degree | 0.03 | 0.03 | 0.04 | 0.08 |
| | College degree | 0.01 | 0.02 | 0.02 | 0.02 |

5. **Conclusions**

This paper introduces and summarizes a new longitudinal data series of the individual-level probability of *estimated* food security estimates from 1979-2019 PSID data. PFS is a proxy for the official household food security measure used in the United States but enables the construction of food security estimates from much longer panel survey data series than has been feasible previously. This data series enables researchers to *estimate* food insecurity when the official food security measure is absent, as well as to explore how longer-run food security dynamics are linked to various policies, shocks, etc. We believe the PFS can contribute, in particular, to longitudinal analysis. Static food insecurity rates estimated by the PFS can differ – especially at the individual level – from those measured by the CPS-FSS, most especially among subpopulations most commonly found to be food insecure by either measure.

The basic descriptive patterns evident in the data are revealing. *Estimated* food insecurity spells are roughly equally likely to be transitory or persistent. Business cycle effects are strongly associated with both individuals starting *estimated* transitory food insecurity spells and longer duration spells among those already *estimated to be* food insecure. Given the conditional persistence of *estimated* food insecurity, at any moment in time most of the food insecure suffer from recurring or chronic *estimated* food insecurity. The subpopulations likely to persist in *estimated* food insecurity are women, non-White, physically disabled, or less well-educated than the overall population. The burden of persistent *estimated* food security falls on a minority of relatively disadvantaged US residents. These patterns echo those found in the prior literature, but the additional insights afforded by the ability to study longer-term food security dynamics only serves to underscore the policy importance of identifying effective interventions to relieve persistent, structural food insecurity among US residents. To date, data constraints have limited researchers' ability to do policy analysis related to food security dynamics longer than the three successive PSID rounds. Our hope is that this new data product, and the PFS method that enables it, can help stimulate further policy research to tackle these and related questions.

# Appendix

## A. Sample construction

### A1. Longitudinal linkage of household data via PSID sample individuals

**Table A1: Longitudinal Linkage of Sample individuals**

| Person ID (PID)/Year | 1968 | … | 1977 | 1978 | 1979 | 1980 | 1981 | 1982 | PSID sample | Study sample |
|---|---|---|---|---|---|---|---|---|---|---|
| 1 | RP-10 | .. | RP-20 | RP-30 | RP-40 | RP-50 | RP-60 | RP-70 | Y | Y |
| 2 | SP-10 | .. | SP-20 | SP-30 | SP-40 | SP-50 | SP-60 | SP-70 | Y | Y |
| 3 | CH-10 | .. | RP-21 | RP-31 | RP-41 | RP-51 | RP-61 | RP-71 | Y | Y |
| 6 | CH-10 | .. | SP-24 | SP-34 | RP-43 | RP-53 | SP-63 | SP-73 | Y | Y |
| 170 | x | | RP-24 | RP-34 | x | x | x | x | N | N |
| 8 | CH-10 | .. | CH-20 | CH-30 | CH-40 | RP-54 | RP-64 | RP-74 | Y | Y |
| 37 | x | | x | x | x | x | CH-64 | CH-74 | Y | N |
| 30 | CH-10 | .. | CH-20 | CH-30 | x | x | x | x | Y | N |
| 100 | RP-11 | .. | x | x | x | x | x | x | Y | N |

PSID collects household data. But there are several methods one can use to construct household dynasties, i.e., the sequences of household-specific observations. One, which LBH followed, is to only include households whose RP does not change over time.

Table A1 shows an example of actual individuals and their roles in their families in the PSID data, with a few minor modifications to illustrate different patterns in a single table, of how individuals change their status in their households over years. We use this to illustrate which individuals we included in our study sample and how we construct household dynasties based on individuals from original (1968) PSID households. Each cell show has the information of the person's role in the household (RP - reference person, SP - spouse, CH - child, x-not surveyed), followed by the PSID year-specific two-digit household ID (HID). The rightmost, "Study sample" column shows whether a person is included in the study sample. The inclusion criterion for our data set is that the individual was in the PSID sample (second column from the right) and was RP or SP at least once during the study period. The "PSID sample" column (second from right) shows whether a person is either initially surveyed in 1968 or his/her lineal descendant (i.e., child, grandchild, etc.). Individuals not in the PSID sample have zero individual PSID survey weights and are therefore excluded from this study sample.

Consider the following example. In 1968, seven individuals were surveyed from two households; a household (HID= 10) of RP (PID=1), spouse (PID=2) and four children (PIDs = 3,6,8,30) and another household (HID=02) of RP only (PID=100). Sometime between 1968 and 1977 – the first year of PSID data we use to estimate PFS to construct the new PFS series – two children from the first household (PID=3,6) had split-off and formed their own households as RP (PID=3) or spouse (PID=6) to a new RP (PID=170), who was not a member (or lineal descendant) of a 1968 PSID sample household. Meanwhile, the RP of the second household (PID=100) disappeared from the PSID sample. In 1979, one of the children who had started their own household by 1977 (PID=6) divorced and became RP of her own household (HID=43), while one of her siblings (PID=30) died. In 1980, the fourth child (PID=8) split-off to form a new household that added a child (PID=37) in the following year.

We use individuals to link together household-level observations over time into a household dynasty defined by those who were (i) part of the original 1968 PSID sample, and either (ii-1) RP or SP in 1977 or (ii-2) child of (ii-1) who eventually became RP or SP in later years. This provides a complete accounting of all households that either continue or originate from the original, nationally representative 1968 PSID sample and are present during the 40-year period we study.

So in this example, our study sample excludes those not in the PSID sample (PID=170), as they have zero individual PSID survey weights, any person who was not surveyed during the study period (PID=100), and those who never were RP or spouse (PID=37, 30). Everyone else is included in our study sample, using their individual weight and the year-specific data from the households of which they were a part. This implies that in periods when multiple individuals in our sample co-resided, that household's data is represented by multiple individuals, but with weights adjusted accordingly. This way we can create a longitudinal sample that tracks the original 1968 households and their lineage comprehensively using household-year-level observations of food expenditures and the household demographic data that determine the household's TFP cost for that year, while appropriately adjusting the weighting of observations of the same household in the same year.

**A2. Construction of Key Variables**

Key variables in our study include (1) SNAP status (2) SNAP benefit amount and (3) food expenditures. Since the way PSID collected these variables has changed over the study period, we carefully constructed these variables to ensure consistency over the period, as we describe here. Table A2 explains the construction of other variables such as household characteristics.

(1) SNAP status (in the previous month)

   From 1977 to 1993, the PSID data does not have the variable that directly observes household SNAP status. Instead, it has the variable how many people in the household received SNAP in the previous month.[13] Based on this variable, we constructed household SNAP status such that the household received SNAP if at least one person received SNAP, and zero otherwise.

   From 1994 to 1997, and from 2009 to 2019, the PSID data has the variable describing whether the household received SNAP in the previous month. We determined household SNAP status such that it received SNAP if answered "yes" and did not receive SNAP if answered otherwise ("no", "N/A", "don't know", etc.)

   From 1999 to 2007, the PSID has the set of 12 dummy variables describing household SNAP status in each month. We constructed household SNAP status such that the household received SNAP if the variable describing whether the household received SNAP one month before the month surveyed.

(2) SNAP benefit amount

   Households reported the SNAP benefit amount they received with different recall periods. We harmonized the benefit amount into monthly amounts. For those who answered as "other/don't know/not applicable/refused to answer" (accounts for less than 2 ppt of the total responses), we used the monthly average value for the same year. We also replaced the zero value as missing if households did not receive SNAP.

(3) (Monthly) Food expenditures

   There are three key features of how PSID collects food expenditures data. First, PSID decomposes food expenditures into three categories: at-home, delivered, and eaten out. Second, since 1994, PSID first asks households whether they received SNAP or not. Respondents are then asked food expenditures if they did not receive SNAP. SNAP recipients, however, are asked the SNAP benefit amount as well as food expenditure <u>in addition to</u> SNAP benefit amount received. Third, food expenditures do NOT include the SNAP benefit amount while we want SNAP benefits included in total household food expenditures.

   From 1977 to 1993, PSID data includes the <u>annual</u> food expenditures of (1) at home and delivered combined and (2) eaten out. We imputed monthly food expenditure by dividing annual expenditures by 12, total monthly food expenditure by summing up those two variables, and total food expenditure including SNAP benefit amount by adding monthly SNAP benefit amount.

---

[13] Although the PSID variable label wrote it as "in the previous year", it is actually the status in the previous month. The author confirmed it from the PSID staff.

From 1994 to 2019, we imputed SNAP-included at-home expenditures by (1) adding up SNAP benefit amount and extra amount spent for SNAP participants, and (2) simply using the reported at-home expenditures for non-SNAP participants. We then added delivered expenditures and eaten-out expenditures to impute total monthly food expenditures including SNAP. For households with different recall periods and other responses, we applied the same procedure as we did with SNAP benefit amount.

Figure A1 shows that the imputed per capita food expenditure is smooth over time. There's a slight change in trend in 1994 when the PSID changed the way food expenditure is collected, but it is still generally smooth over the years. We validated our imputed food expenditure by comparing it with the total household food expenditures included in the raw data since 1999. We find that the difference in total monthly expenditures is $11.2 on average (2% of mean food expenditure), and 95% of the differences are less than $6 (1% of mean food expenditure). These small differences imply that our imputed expenditures are pretty accurate.

**A3. Imputation of missing individual-level race and educational attainment**

The PSID does not collect an individual's race if a person is neither RP (entire study-period) nor SP (since 1985). So, we simply impute the value using the individual's race from the survey round when the individual's race was first collected. After this imputation, 1.6% of the observations (4,293 observations from 601 individuals) have missing race information over the entire study period because they were only a SP pre-1985. PSID likewise does not collect educational attainment if a person is under 16 years old. In the case of children 15 and younger, we use the RP's educational attainment as that is more salient to the household's food access status. This leaves just 0.9% of total observations (2,435 observations) with missing educational information.

**A4. Nationally representative sample**

Compared to official US household estimates based on the Current Population Survey (CPS) (U.S. Census Bureau. 2022), our sample's household characteristics are similar in composition and show similar trends over the study period, implying that the sample maintains reasonably national representativeness. Figure A2 shows the sex and racial composition of RPs in the study sample and overall, per Census estimates. Our sample underestimates the share of female-RP households, possibly due to the nature of the PSID; the PSID tags a male partner as a RP in a household with a married couple of different sexes. The difference in race between our sample and the Census estimate is very small. Most importantly, both series show similar trends over time.

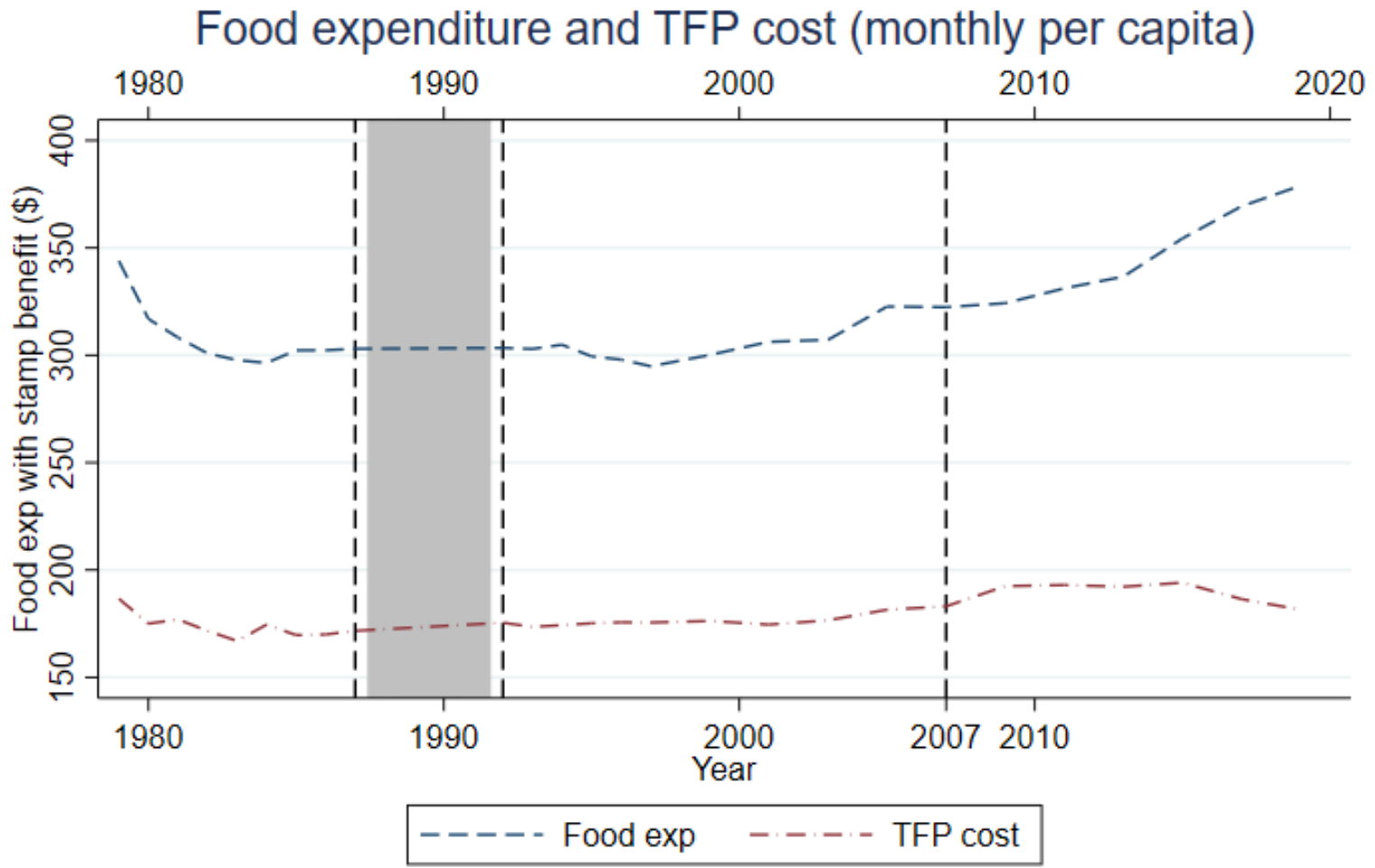

Figure A1: Monthly per capita food expenditure and TFP cost over years.

**Table A2. Variable Description**

| Variable | Note |
|---|---|
| Marital status (RP) | Single, widowed, separated, and divorced are categorized as "NOT married" |
| Race (RP) | Only counts the first response of the individuals with multiple races. All races that are NOT White (Black, Asian, Native American, etc.) are categorized as "non-White" |
| Region of residence (RP) | Northeast: ME, NH, VT, NY, MA, CT, RI<br><br>Mid-Atlantic: PA, NJ, DC, DE, MD, VA<br><br>South: NC, SC, GA, TN, WV, FL, AL, AR, MS, LS, TX<br><br>Midwest: OH, IN, MI, IL, MN, WI, IA, MO<br><br>West: KS, NE, ND, SD, OK, AZ, CO, ID, MT, NV, NM, UT, WY, OR, WA, CA |
| Employment (RP) | Employed = "Working now" and "Temporarily laid off"<br><br>NOT employed = "Looking for work", "retired", "permanently disabled", "housekeeping", "student", "other"<br><br>Disabled if has limitation to either the type of work or the amount of work. |
| Education (RP) | Less than high school: Less than 12 grades completed.<br><br>High school: Completed 12 grades or has GED.<br><br>Some college: "College, but no degree", or completed 13 grades or more but answered "No" to the question "Has college degree."<br><br>College: Answered "Yes" to the question "Has college degree" |
| Disability (RP) | Disabled if has limitation to either the type of work or the amount of work |

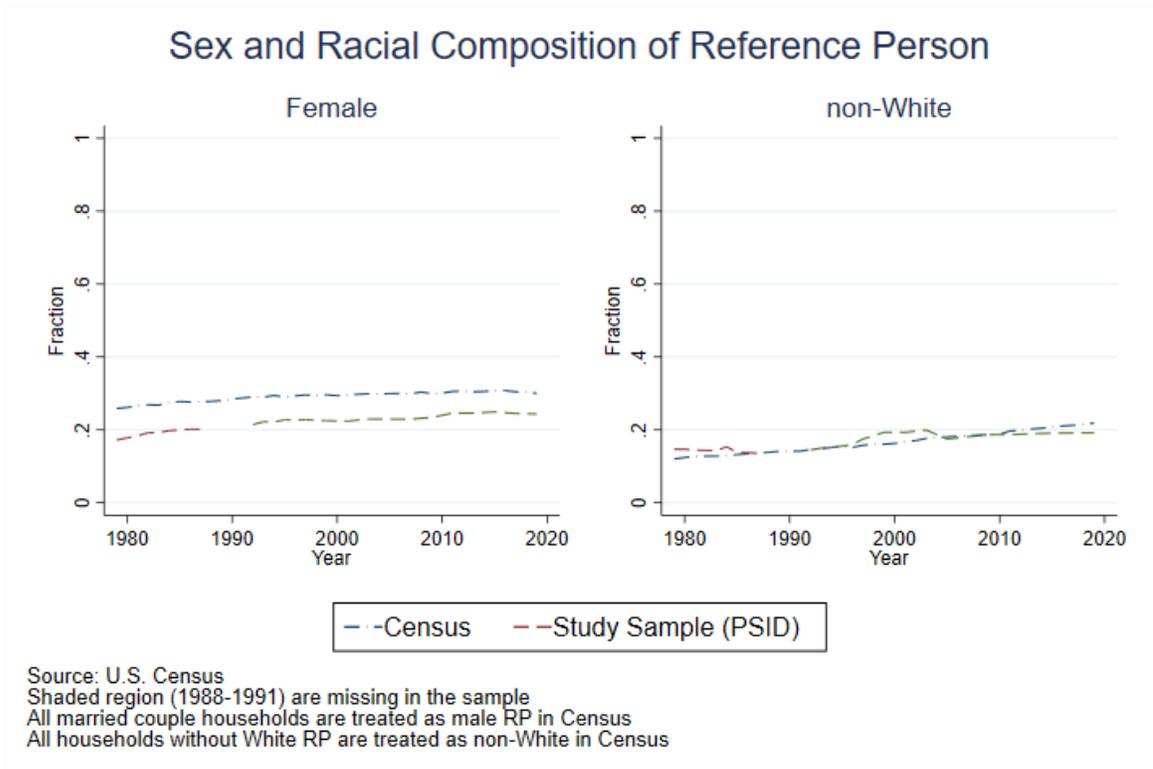

**Figure A2**: Sex and Racial Composition of Reference Person in the sample and Census

## B. Additional Tables and Figures

### Table B1: Conditional Mean and Variance of Food Expenditure per capita

|  | MLE | | Marginal Effects | |
|---|---|---|---|---|
|  | Food exp per capita | Var (food exp) | Food exp per capita | Var (food exp) |
| Food exp 2 years ago | 0.00*** | 0.00*** | 0.40*** | 37.76*** |
|  | (0.00) | (0.00) | (0.01) | (5.32) |
| (Food exp 2 years ago)$^2$ (K) | -0.00*** | -0.00*** | -0.10*** | -7.35*** |
|  | (0.00) | (0.00) | (0.01) | (1.59) |
| Age (RP) | 0.01*** | -0.03*** | 1.92*** | -525.94*** |
|  | (0.00) | (0.00) | (0.18) | (44.44) |
| (Age (RP))$^2$/1000 | -0.06*** | 0.24*** | -18.33*** | 4259.90*** |
|  | (0.01) | (0.02) | (1.78) | (441.52) |
| non-White (RP) | -0.05*** | 0.14*** | -16.55*** | 2423.97*** |
|  | (0.01) | (0.02) | (1.78) | (400.41) |
| Married | -0.10*** | -0.42*** | -30.42*** | -7414.80*** |
|  | (0.01) | (0.02) | (1.86) | (446.83) |
| Female (RP) | -0.11*** | -0.23*** | -35.27*** | -4045.54*** |
|  | (0.01) | (0.02) | (2.02) | (413.30) |
| Less than High School | -0.01*** | 0.04 | -4.61*** | 625.20 |
|  | (0.00) | (0.02) | (1.57) | (410.87) |
| College (w/o degree) | 0.04*** | 0.08*** | 12.04*** | 1431.47*** |
|  | (0.00) | (0.03) | (1.54) | (475.83) |
| College Degree | 0.04*** | 0.06** | 14.10*** | 1049.88** |
|  | (0.00) | (0.03) | (1.48) | (524.09) |
| Employed | 0.03*** | 0.04* | 10.14*** | 696.10* |
|  | (0.00) | (0.02) | (1.37) | (396.57) |
| Disabled | -0.01*** | 0.04** | -4.71*** | 748.72** |
|  | (0.00) | (0.02) | (1.28) | (368.34) |
| FU size | -0.08*** | -0.22*** | -25.51*** | -4000.84*** |
|  | (0.00) | (0.01) | (0.57) | (215.15) |
| % of children | -0.12*** | -0.59*** | -38.33*** | -1.1e+04*** |
|  | (0.01) | (0.05) | (2.79) | (924.99) |
| RP changed | 0.10*** | -0.03** | 32.40*** | -565.12** |
|  | (0.00) | (0.01) | (1.07) | (250.14) |
| ln(per capita income) | -0.08*** | -0.39*** | -25.01*** | -6928.62*** |
|  | (0.01) | (0.04) | (2.04) | (688.79) |
| Received SNAP | 0.00*** | 0.00*** | 0.40*** | 37.76*** |
|  | (0.00) | (0.00) | (0.01) | (5.32) |
| Constant | 4.49*** | 11.08*** |  |  |
|  | (0.03) | (0.14) |  |  |
| N | 270,303 | 270,303 | 270,303 | 270,303 |

Table B2: Association of PFS with individual and household characteristics

|  | PFS | PFS | PFS | PFS |
|---|---|---|---|---|
| Age (RP) | 0.007*** | 0.007*** | 0.006*** | 0.006*** |
|  | (0.00) | (0.00) | (0.00) | (0.00) |
| Age squared (RP)/1000 | -0.060*** | -0.061*** | -0.062*** | -0.061*** |
|  | (0.00) | (0.00) | (0.00) | (0.00) |
| non-White (RP) | -0.078*** | -0.078*** | -0.058*** | -0.057*** |
|  | (0.00) | (0.00) | (0.00) | (0.00) |
| Married (RP) | 0.047*** | 0.046*** | 0.044*** | 0.045*** |
|  | (0.00) | (0.00) | (0.00) | (0.00) |
| Female (RP) | -0.024*** | -0.025*** | -0.024*** | -0.023*** |
|  | (0.00) | (0.00) | (0.00) | (0.00) |
| Less than High School (RP) | -0.032*** | -0.024*** | -0.026*** | -0.018*** |
|  | (0.00) | (0.00) | (0.00) | (0.00) |
| College (w/o degree) (RP) | 0.031*** | 0.032*** | 0.025*** | 0.026*** |
|  | (0.00) | (0.00) | (0.00) | (0.00) |
| College (RP) | 0.035*** | 0.036*** | 0.031*** | 0.032*** |
|  | (0.00) | (0.00) | (0.00) | (0.00) |
| Employed (RP) | 0.019*** | 0.019*** | 0.019*** | 0.019*** |
|  | (0.00) | (0.00) | (0.00) | (0.00) |
| Disabled (RP) | -0.023*** | -0.022*** | -0.021*** | -0.021*** |
|  | (0.00) | (0.00) | (0.00) | (0.00) |
| HH size | -0.044*** | -0.043*** | -0.039*** | -0.040*** |
|  | (0.00) | (0.00) | (0.00) | (0.00) |
| % of children | 0.073*** | 0.075*** | 0.053*** | 0.055*** |
|  | (0.00) | (0.00) | (0.00) | (0.00) |
| RP changed | -0.014*** | -0.014*** | -0.015*** | -0.015*** |
|  | (0.00) | (0.00) | (0.00) | (0.00) |
| ln(per capita income) | 0.082*** | 0.082*** | 0.073*** | 0.073*** |
|  | (0.00) | (0.00) | (0.00) | (0.00) |
| Received SNAP | -0.068*** | -0.068*** | -0.052*** | -0.052*** |
|  | (0.00) | (0.00) | (0.00) | (0.00) |
| Constant | -0.111*** | -0.114*** |  | -0.116*** |
|  | (0.01) | (0.01) | -0.000 (0.01) | (0.03) |
| N | 270303 | 267864 | 270025 | 267580 |
| $R^2$ | 0.77 | 0.77 | 0.83 | 0.83 |
| Individual-level controls | N | Y | N | Y |
| Individual FE | N | N | Y | Y |

Note: Individual controls include age, age squared, and educational attainment. State and year fixed effects are included in all specifications.

**Table B3: Correlation Coefficients among PFS Thresholds and Macroeconomic Indicators**

| Variables | (1) | (2) | (3) | (4) | (5) | (6) |
|---|---|---|---|---|---|---|
| (1) PFS Cut-off | 1.000 | | | | | |
| (2) ln (disposable personal income per capita) | 0.818*** | 1.000 | | | | |
| (3) SNAP participation rate | 0.540** | 0.541** | 1.000 | | | |
| (4) Poverty rate | -0.290 | -0.205 | 0.570** | 1.000 | | |
| (5) Unemployment Rate | -0.095 | 0.069 | 0.465* | 0.822*** | 1.000 | |
| (6) GDP growth per capita | -0.165 | -0.282 | -0.190 | -0.307 | -0.662*** | 1.000 |

*** $p<0.01$, ** $p<0.05$, * $p<0.1$

**Table B4: Estimated Food Security Status as estimated by PFS and FSSS - Reclassified**

| Status measured by PFS / Status measured my FSSS | 1999 | 2001 | 2003 | 2015 | 2017 | 2019 | Total |
|---|---|---|---|---|---|---|---|
| Food secure / Food secure | 0.82 | 0.81 | 0.79 | 0.80 | 0.80 | 0.82 | 0.81 |
| Food insecure / Food insecure | 0.04 | 0.05 | 0.05 | 0.04 | 0.04 | 0.03 | 0.04 |
| Food insecure / Food secure | 0.08 | 0.07 | 0.08 | 0.10 | 0.08 | 0.08 | 0.08 |
| Food secure / Food insecure | 0.06 | 0.07 | 0.08 | 0.07 | 0.08 | 0.07 | 0.07 |

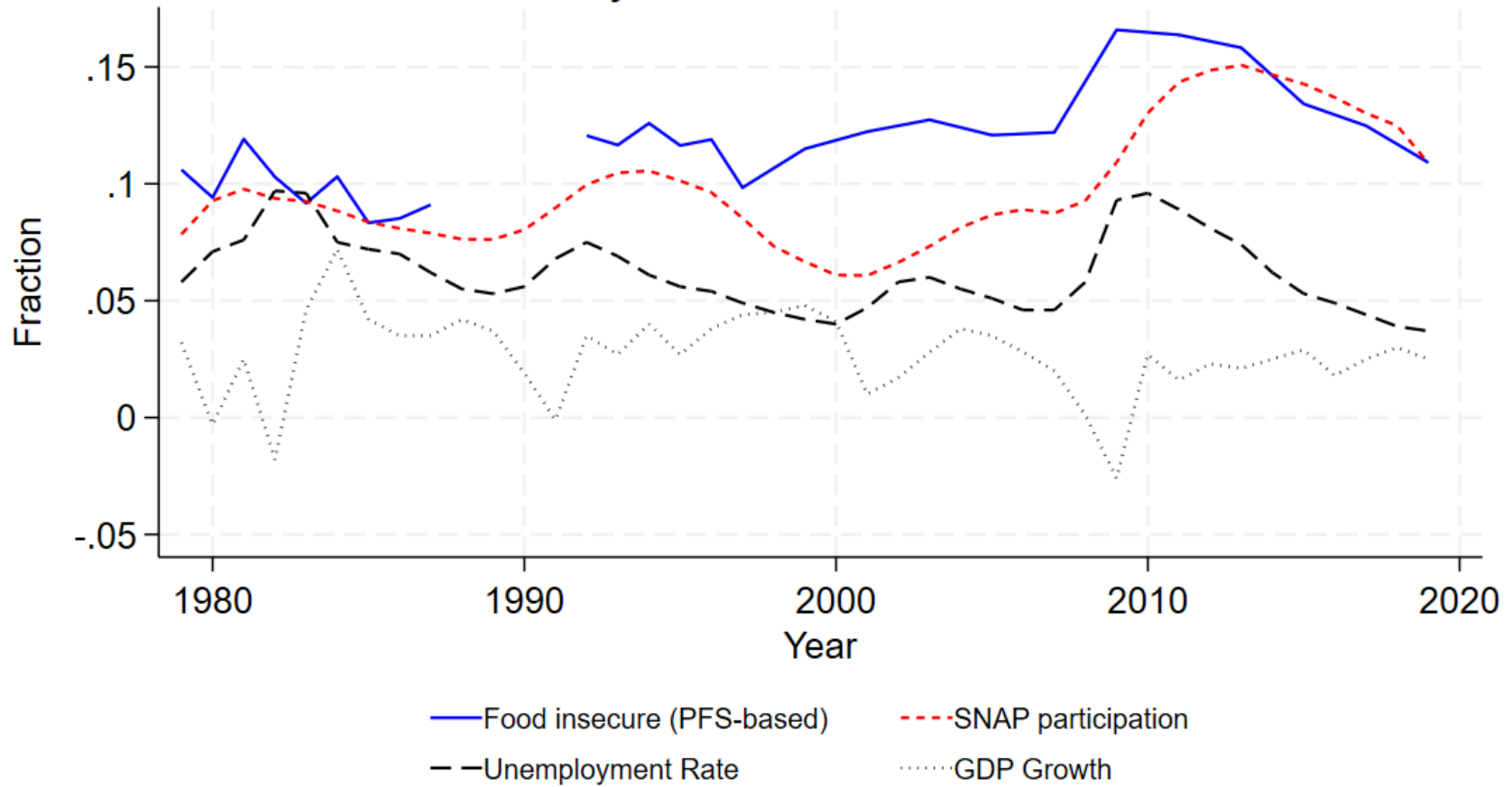

Figure B1: Estimated Food Insecurity, SNAP participation, unemployment and GDP Growth rates

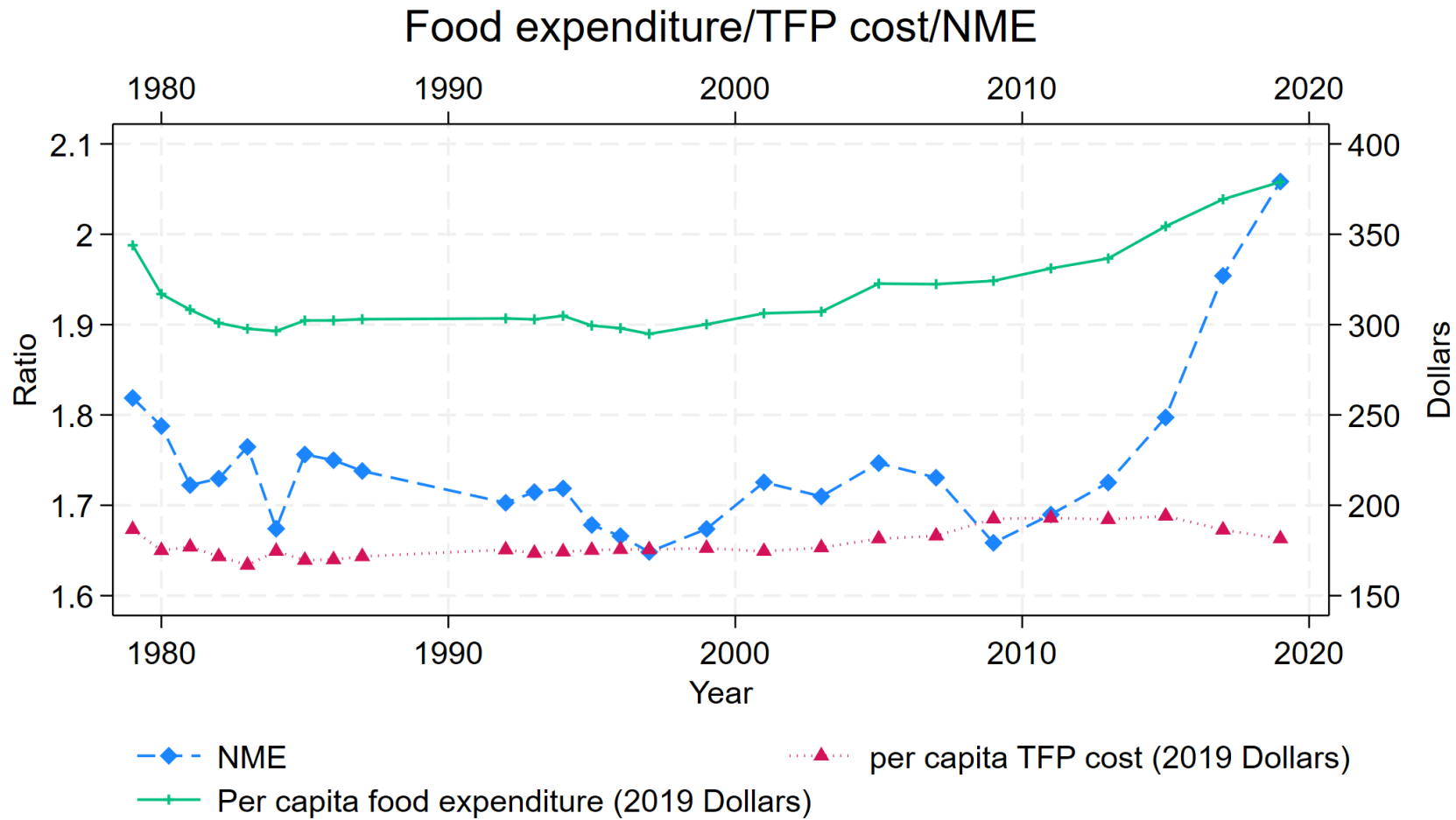

Figure B2: Food expenditure, TFP cost and Normalized Monetary Expenditure